\documentclass[12pt]{iopart}
\usepackage{graphicx,graphics,color,epsfig}
\begin{document}
\title{Shot noise of spin current and spin transfer torque}
\author{
Yunjin Yu$^1$, Hongxin Zhan$^1$, Langhui Wan$^1$, Bin Wang$^{1,2}$, Yadong Wei$^1$,Qingfeng Sun,$^2$ and Jian Wang$^3$
}
\address{$^1$ College of Physics Science and Technology and Institute of
Computational Condensed Matter Physics,
Shenzhen  University, Shenzhen 518060, China}
\address{$^2$ Institute of Physics, Chinese Academy of Sciences, Beijing, China}
\address{$^3$ Department of Physics and The Center of Theoretical and
Computational Physics, The University of Hong Kong, China}
%\ead{}

\begin{abstract}
We report the theoretical investigation of noise spectrum of spin current ($S^\sigma$)
and spin transfer torque ($S^\tau$) for non-colinear spin polarized transport in a
spin-valve device which consists of normal scattering region connected by two
ferromagnetic electrodes (MNM system). Our theory was developed using non-equilibrium
Green's function method and general non-linear $S^\sigma-V$ and $S^\tau-V$ relations
were derived as a function of angle $\theta$ between magnetization of two leads. We
have applied our theory to a quantum dot system with a resonant level coupled with
two ferromagnetic electrodes. It was found that for the MNM system, the auto-correlation
of spin current is enough to characterize the fluctuation of spin current. For a system
with three ferromagnetic layers, however, both auto-correlation and cross-correlation of
spin current are needed to characterize the noise spectrum of spin current.
Furthermore, the spin transfer torque and the torque noise were studied for the MNM system.
For a quantum dot with a resonant level, the derivative of spin torque with respect to bias
voltage is proportional to $\sin\theta$ when the system is far away from the resonance.
When the system is near the resonance, the spin transfer torque becomes non-sinusoidal
function of $\theta$. The derivative of noise spectrum of spin transfer torque with
respect to the bias voltage $N_\tau$ behaves differently when the system is near or
far away from the resonance. Specifically, the differential shot noise of spin transfer
torque $N_\tau$ is a concave function of $\theta$ near the resonance while it becomes
convex function of $\theta$ far away from resonance. For certain bias voltages, the
period $N_\tau(\theta)$ becomes $\pi$ instead of $2\pi$. For small $\theta$, it was
found that the differential shot noise of spin transfer torque is very sensitive
to the bias voltage and the other system parameters.
\end{abstract}

\pacs{72.25.-b,72.70.+m,74.40.-n}
\submitto{Nanotechnology}

\section{Introduction}
Electronic shot noise describes the fluctuation of current and is an
intrinsic property of quantum devices due to the quantization of electron
charge. In the past decade, the study of shot noise has attracted increasing
attention\cite{BBreviewB} because it can give additional information that
is not contained in the conductance or charge current. It can be used to probe
the kinetics of electron\cite{RLandauer} and investigate correlations of electronic
wave functions\cite{TGramespacherandMButtiker}. In the study of shot noise
$<(\Delta\hat{I})^2>$, the Fano Factor $F=<(\Delta\hat{I})^2>/2q<\hat{I}>$ is often used
where $<\hat{I}>$ is the current. When $F>1$ it is referred as super-Poissonian noise,
while $F<1$ corresponds to sub-Poissonian behavior. In general, for a quantum device,
Pauli exclusion suppresses the shot noise and hence reduces the Fano
factor\cite{VAKhlus,MButtikerPRL,MButtikerPRB} but Coulomb interaction can
either suppress or enhance shot noise depending on system
details\cite{TGonzalez,CWBeenakker,GIannacconeandGLombardi,VVKuznetsovandEEMendez,QChen}.
The suppression of shot noise has been confirmed experimentally in quantum point
contact\cite{MReznikovandMHeiblum,LDiCarloandYZhang}, single electron tunneling
regime\cite{HBirkandMJMdeJong,ANauenandIHapkeWurst}, graphene
nano-ribbon\cite{RDanneauandFWu,RDanneauandFWu2}, and atom-size
metallic contacts\cite{HEvandenBromandJMvanRuitenbeek,BinWang}. The enhancement
of shot noise was also observed in GaAs based quantum contacts when the system is in
the negative differential conductance region\cite{SSSafonovandAKSavchenko}.
Recently, with the development of spintronics, polarized spin current especially
pure spin current received much more attention. Less attention has been paid on
the polarized spin current correlation compared with the charge current correlation\cite{YCChenandMDiVentra,Baigengwang2004,YCChenandMDiVentra2,HKZhaoandLLZhao,BinwangPRB,HKZhaoandLLZhao2}.
Shot noise of polarized spin current has been studied in several quantum devices including the
MNM(ferromagnet-normal-ferromagnet)\cite{HZhaoandJianWang} and NMN(normal-magnetic-normal)\cite{Ouyang}.
In these devices, shot noise is expected to provide additional information about the spin-dependent
scattering process and spin accumulation. It was shown that shot noise can be
used to probe attractive or repulsive interactions in mesoscopic systems and
to measure the spin relaxation time~\cite{sauretOandFeinberg}.
For a two-probe normal system (NNN system), it is well known that the charge current correlation between
different probes (cross correlation noise) is negative definitely\cite{SanchezD},
but for a magnetic junction, the spin cross correlation noise between
different probes is not necessarily negative due to
spin flip mechanism. For example, Ref.\cite{MinhuiShangguan} showed that the cross
correlation can be positive at special Fermi energy due to Rashba interaction.

Recently, spin transfer torque (STT), predicted by Slonczewski\cite{Slonczewski1,Slonczewski2}
and Berger\cite{Berger}, has been the subject of intensive investigations\cite{Tserkovnyak,Xiake,Haney,Mahfouzi}.
Spin current can transfer spin angular momentum and be used to
switch the magnetic orientation of ferromagnetic layers in GMR and TMR devices.
Therefore, STT has potential applications\cite{Katine} such as
hard-disk read head\cite{Takagishi}, magnetic detection sensor\cite{Braganca},
and random access memory (MRAM)\cite{SUNZY}, etc..
It comes from the absorption of the itinerant flow of angular momentum
components normal to the magnetization direction and relies on the
system spin polarized current. The noise spectrum of STT drastically affects
the magneto-resistance behavior\cite{Chudnovskiy}. Many studies have focused
on the spin transfer torque in various materials and under dc or ac condition.
The correlation effect or quantum noise of spin transfer torque has not been
studied so far. It is the purpose of this paper to fill this gap. In this paper,
we have calculated the shot noise of particle current, spin current as well as
spin transfer torque in the nonlinear regime for a magnetic quantum dot connected
with two non-colinear magnetic electrodes. We found that for a MNM spin-valve
system, the spin auto-correlation is enough to characterize the fluctuation of
spin current. For a system with three ferromagnetic layers (MNMNM), however,
both auto-correlation and cross-correlation are needed to characterize the
fluctuation of spin current. For the quantum dot with a resonant level, the
behavior of differential STT depends on whether the system on resonance or
off resonance. When the system is off resonance, the differential STT reduces
to the familiar result of tunneling barrier $(1/2)(I^s(\pi)-I^s(0)) \sin\theta$
where $\theta$ is the angle between magnetic moments of ferromagnetic leads.
If it is on resonance, the dependence of differential STT on $\theta$ becomes
non-sinusoidal. The resonance also has influence on the noise spectrum of STT.
If the system is near the resonance, noise spectrum of STT is a concave function
of $\theta$ while it becomes a convex function far away from the resonance.

This paper is organized as follows. Firstly, we derive the general formulae of
spin auto-correlation shot noise, spin cross-correlation shot noise and spin transfer
torque shot noise from the non-equilibrium Green's function method. Then we analyze
the spin transport properties for the MNM system. Finally, we give the conclusions.

\section{Theory formalism}
We start from the Hamiltonian of the quantum dot which is connected by
two magnetic electrodes. We assume that the current flows in the $y(y^{'})$-direction
and the left lead magnetic moment ${\bf M}_L$ always points at
the $z$-direction, while the right lead magnetic moment ${\bf M}_R$ points at
an angle $\theta_R$ to the $z$-direction in $x-z$ plane (see figure 1).

\begin{figure}
\includegraphics[height=6cm,width=8cm]{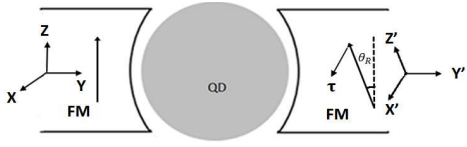}
\vspace{-2.0cm}
\caption{The schematic plot of the spin-degenerated quantum dot connected
by two ferromagnetic leads. The magnetic moment of the left lead is always at
$z$ direction and the magnetic moment of the right lead is at an
angle $\theta_{R}$ with respect to the $z$-axis in the $x-z$ plane.}
\label{Fig.1}
\end{figure}

In the second quantized form, Hamiltonian is
\begin{equation}
\hat{H}=\hat{H}_{lead}+\hat{H}_{dot}+\hat{H}_{T} ,
\label{Ham}
\end{equation}
where $\hat{H}_{lead}$ is the Hamiltonian of the leads,
\begin{eqnarray}
\hat{H}_{lead}=\sum_{k_\alpha\sigma}(\epsilon_{k_\alpha}-\sigma M_{\alpha} cos\theta_\alpha)
\hat{C}^{\dagger}_{k_\alpha\sigma}\hat{C}_{k_\alpha\sigma}
-\sum_{k_\alpha\sigma} M_{\alpha} sin\theta_\alpha \hat{C}^{\dagger}_{k_\alpha\sigma}\hat{C}_{k_\alpha \bar \sigma},
\label{Hamlead}
\end{eqnarray}
where $\hat{C}^{\dagger}_{k_\alpha\sigma}$  creates an electron in lead $\alpha$ with
energy level $k$ and spin $\sigma$, $\sigma=\pm 1$ and $\bar \sigma=-\sigma$.
The second term $\hat{H}_{dot}$ is the Hamiltonian of the isolated quantum dot,
\begin{equation}
\hat{H}_{dot}=\sum_{n\sigma} \epsilon_{n}\hat{d}^{\dagger}_{n\sigma}\hat{d}_{n\sigma},
\label{Hamdot}
\end{equation}
The third term $\hat{H}_{T}$ is the Hamiltonian describing the coupling between quantum
dot and the leads with the coupling constant $t_{k_\alpha n\sigma}$,
\begin{equation}
\hat{H}_{T}=\sum_{k_\alpha n\sigma}[t_{k_\alpha n\sigma}\hat{C}^{\dagger}_{k_\alpha\sigma}\hat{d}_{n\sigma}+c.c.] ,
\label{HamT}
\end{equation}
where $c.c.$ denotes the complex conjugate. By applying the Bogoliubov transformation~\cite{Bogoliubov},
\begin{equation}
\hat{c}_{k_\alpha\sigma}=cos(\theta_\alpha/2)\hat{C}_{k_\alpha\sigma}-\sigma sin(\theta_\alpha/2)\hat{C}_{k_\alpha\bar\sigma},
\label{Bogo}
\end{equation}
we can diagonalize the Hamiltonian of the electrodes to give the following effective Hamiltonian,
\begin{equation}
\hat{H}_{\alpha}=\sum_{k\sigma}(\epsilon_{k_\alpha}-\sigma M_\alpha)\hat{c}^{\dagger}_{k_\alpha\sigma}\hat{c}_{k_\alpha\sigma}.
\label{Hamleadnew}
\end{equation}
So for a ferromagnetic lead coupled with scattering quantum dot,
the line-width function $\Gamma_{\alpha}$ can be written as~\cite{B.G.Wang}
\begin{eqnarray}
\Gamma_{\alpha}=R_{\alpha}\left(
  \begin{array}{ccc}
    \Gamma_{\alpha\uparrow} & 0 \\
    0 & \Gamma_{\alpha\downarrow}\\
  \end{array}
\right)R^{\dagger}_{\alpha},
\end{eqnarray}
where
 \begin{equation}
R_{\alpha}=\left(
  \begin{array}{ccc}
    cos\frac{\theta_{\alpha}}{2} & -sin\frac{\theta_{\alpha}}{2} \\
    sin\frac{\theta_{\alpha}}{2} & cos\frac{\theta_{\alpha}}{2}\\
  \end{array}
\right)
\end{equation}
is the rotational matrix.

The current operator of the lead $\alpha$ with spin $\sigma$ is defined as
\begin{equation}
\hat{I}_{\alpha\sigma}(t)=q\frac{d\hat{N}_{\alpha\sigma}}{dt}.
\label{CurOperator}
\end{equation}
where $\hat{N}_{\alpha\sigma}=\sum_{k}\hat{C}^{\dagger}_{k_\alpha\sigma}\hat{C}_{k_\alpha\sigma}$
is the number operator for the electron in the lead $\alpha$.
By using the Heisenberg equation of motion, we have
\begin{equation}
\hat{I}_{\alpha\sigma}(t)=-i\frac{q}{\hbar}\sum_{km}[t_{k_\alpha m \sigma}
\hat{C}_{k_\alpha\sigma}^{\dagger}(t)\hat{d}_{m\sigma}(t)]+c.c.
\label{CurOperator}
\end{equation}
The average current can be expressed by in terms of Green's function,
\begin{equation}
<\hat{I}_{\alpha\sigma}(t)>=-\frac{q}{\hbar}\sum_{km}[t_{k_\alpha m\sigma}G^{<}_{m\sigma k_\alpha\sigma}(t,t)+c.c].
\label{CurAverage}
\end{equation}
If we consider the total charge current flowing through the lead $\alpha$,
the charge current operator can be expressed as
\begin{equation}
\hat{I}_{\alpha}=\hat{I}_{\alpha\uparrow}+\hat{I}_{\alpha\downarrow}
\end{equation}
and the spin current operator in $z$-direction is
\begin{equation}
\hat{I}^{s}_{\alpha}=\frac{\hbar}{2}\frac{1}{q}(\hat{I}_{\alpha\uparrow}-\hat{I}_{\alpha\downarrow}).
\end{equation}
Since the local spin current is not conserved, the loss of the spin angular momentum is transferred to the magnetization of the free layer.
For spin transfer torque, we are interested in the MNM system and we assume that electron coming from the left lead which is pinned and
the right lead is the free layer. The spin transfer torque can be calculated as follows. The total spin of the right ferromagnetic electrode is\cite{Sugang1}

\begin{equation}
 \hat{S}^{\theta}=\frac{\hbar}{2}\sum_{k_R\mu\nu}C_{k_R\mu}^{\dagger}
 C_{k\nu}(\mathcal{R}^{-1}\chi_{\mu})^{\dagger}\hat{\sigma}(\mathcal{R}^{-1}\chi_{\nu}).
 \label{totalspin}
 \end{equation}

Here, $\hat{\sigma}$ is Pauli matrices and the spinup state $\chi_{\mu(\nu)}=(\begin{array}{ccc}
    1\\ 0 \end{array})$ for $\mu(\nu)=1$ or the spindown state $\chi_{\mu(\nu)}=(\begin{array}{ccc}
    0\\ 1 \end{array})$ for $\mu(\nu)=-1$.
Note that the equation above is written in $xyz$ coordinate frame while $\hat{S}^{\theta}$
are quantized in the $x'y'z'$ frame. Because $\hat{S}^{\theta}(t)$ is along the direction of $z'$, the total spin torque
$\hat{\tau}=\frac{\partial \hat{S}^{\theta}}{\partial t}=\frac{i}{\hbar}[\hat{H}_{T},\hat{S}^{\theta}]$
should be along the direction of $x'$ (see figure 1).
So we need the expression of the spin operator of the right lead along $x'$ direction,
$\hat{S}^{\theta}_{x^{'}}$, which can be obtained from equation (\ref{totalspin}),
\begin{eqnarray}
 \hat{S}^{\theta}_{x^{'}}=\frac{\hbar}{2}\sum_{k_{R}\sigma}(\hat{C}^{\dagger}_{k_{R}\sigma}
 \hat{C}_{k_{R}\bar{\sigma}}cos\theta-\sigma\hat{C}^{\dagger}_{k_{R}\sigma}\hat{C}_{k_{R}\sigma}sin\theta).
\end{eqnarray}
According to the Heisenberg
equation of motion, the spin transfer torque operator is
\begin{eqnarray}
\hat{\tau}_{R}&=&\hat{\tau}_{x^{'}}=\frac{i}{\hbar}[\hat{H}_{T},\hat{S}^{\theta}_{x{'}}]\nonumber \\
&=&-\frac{i}{2}[\sum_{k_Rn\sigma\sigma^{'}}(\hat{C}_{k_R\sigma}^{\dagger}
\mathcal{R}_{\sigma\sigma^{'}}\hat{d}_{n\sigma^{'}}t_{k_Rn\sigma^{'}}-t_{k_Rn\sigma^{'}}^{*}
\hat{d}_{n\sigma^{'}}^{\dagger}\mathcal{R}_{\sigma\sigma^{'}}\hat{C}_{k_R\sigma})]
\end{eqnarray}
where,
\begin{equation}
\mathcal{R}=\left(
  \begin{array}{ccc}
    \mathcal{R}_{\uparrow\uparrow} & \mathcal{R}_{\uparrow\downarrow} \\
    \mathcal{R}_{\downarrow\uparrow} & \mathcal{R}_{\downarrow\downarrow}\\
  \end{array}
\right)=\left(
  \begin{array}{ccc}
    -sin\theta_{R} & cos\theta_{R} \\
    cos\theta_{R} & sin\theta_{R}\\
  \end{array}
\right)
,t_{k_{R}n}=\left(
  \begin{array}{ccc}
    t_{k_{R}n\uparrow} & 0 \\
    0 & t_{k_{R}n\downarrow}\\
  \end{array}
\right).
\end{equation}

The average spin transfer torque is\cite{Sugang1}
\begin{eqnarray}
<\hat{\tau}_R>&=&Re\{\sum_{k_Rn} Tr_{\sigma}[t_{k_{R}n}\mathcal{R}G_{n,k_{R}}^{<}]\}\nonumber\\
 &=&\int \frac{dE}{2\pi}(f_{L}-f_{R})Tr[G^{r}\Gamma_{L}G^{a}(i\Sigma_{R}^{a}\mathcal{R}-i\mathcal{R}\Sigma_{R}^{r})]
\label{spintorque}
\end{eqnarray}
where $Tr_\sigma$ is over spin space.

The correlation of the charge current is given by
\begin{eqnarray}
<\Delta\hat{I}_{\alpha}(t_1)\Delta\hat{I}_{\beta}(t_2)>=\sum_{\sigma\sigma{'}}(<\Delta \hat{I}_{\alpha\sigma}(t_1)\Delta \hat{I}_{\beta\sigma{'}}(t_2)>)
\label{curnoise}
\end{eqnarray}
and the shot noise of spin current is
\begin{eqnarray}
<\Delta\hat{I}^{s}_{\alpha}(t_1)\Delta\hat{I}^{s}_{\beta}(t_2)> =\frac{1}{4}\frac{\hbar^{2}}{q^2}\sum_{\sigma\sigma{'}}\sigma\sigma{'}(<\Delta \hat{I}_{\alpha\sigma}(t_1)\Delta \hat{I}_{\beta\sigma{'}}(t_2)>)
\label{spincurnoise}
\end{eqnarray}
where
\begin{eqnarray}
\Delta \hat{I}_{\alpha\sigma}=\hat{I}_{\alpha\sigma}-<\hat{I}_{\alpha\sigma}>,
\end{eqnarray}
and $\sigma=\uparrow\downarrow$ or $\pm 1$.
Finally, the correlation of spin transfer torque is
\begin{eqnarray}
{\cal S}(t_1,t_2)=<\Delta\hat{\tau}_R(t_1)\Delta\hat{\tau}_R(t_2)>,
\label{sttnoise}
\end{eqnarray}
where $\Delta\hat{\tau}_R={\hat \tau}_R - <{\hat \tau_R}>$.

We now derive the correlation of charge current, spin current, and spin transfer torque. Clearly, all correlation
functions contain the following term,
\begin{eqnarray}
&&<\hat{I}_{\alpha\sigma}(t_{1})\hat{I}_{\beta\sigma{'}}(t_{2})>=-\frac{q^{2}}{\hbar^2}\sum_{kk{'}mn}\nonumber\\
&&[t_{k_\alpha m \sigma}t_{k'_\beta n \sigma{'}} <\hat{C}^{\dagger}_{k_\alpha\sigma}(t_{1})\hat{d}_{m\sigma}(t_{1})\hat{C}^{\dagger}_{k'_\beta\sigma{'}}(t_{2})\hat{d}_{n\sigma{'}}(t_2)>\nonumber\\
&&+t^{\ast}_{k_\alpha m \sigma}t^{\ast}_{k'_\beta n \sigma{'}} <\hat{d}^{\dagger}_{m\sigma}(t_{1})\hat{C}_{k_\alpha\sigma}(t_{1})\hat{d}^{\dagger}_{n\sigma{'}}(t_2)\hat{C}_{k'_\beta\sigma{'}}(t_{2})>\nonumber\\
&&-t_{k_\alpha m \sigma}t^{\ast}_{k'_\beta n \sigma{'}} <\hat{C}^{\dagger}_{k_\alpha\sigma}(t_{1})\hat{d}_{m\sigma}(t_{1})\hat{d}^{\dagger}_{n\sigma{'}}(t_2)\hat{C}_{k'_\beta\sigma{'}}(t_{2})>\nonumber\\
&&-t^{\ast}_{k_\alpha m \sigma}t_{k'_\beta n \sigma{'}} <\hat{d}^{\dagger}_{m\sigma}(t_{1})\hat{C}_{k_\alpha\sigma}(t_{1})\hat{C}^{\dagger}_{k'_\beta\sigma{'}}(t_{2})\hat{d}_{n\sigma{'}}(t_2)>]\nonumber,\\
\label{Curtimes}
\end{eqnarray}
and
\begin{eqnarray}
&&<\hat{I}_{\alpha\sigma}(t_{1})><\hat{I}_{\beta\sigma{'}}(t_{2})>=-\frac{q^{2}}{\hbar^2}\sum_{kk{'}mn}\nonumber\\
&&[t_{k_\alpha m \sigma}t_{k'_\beta n \sigma{'}} <\hat{C}^{\dagger}_{k_\alpha\sigma}(t_{1})\hat{d}_{m\sigma}(t_{1})><\hat{C}^{\dagger}_{k'_\beta\sigma{'}}(t_{2})\hat{d}_{n\sigma{'}}(t_2)>\nonumber\\
&&+t^{\ast}_{k_\alpha m \sigma}t^{\ast}_{k'_\beta n \sigma{'}} <\hat{d}^{\dagger}_{m\sigma}(t_{1})\hat{C}_{k_\alpha\sigma}(t_{1})><\hat{d}^{\dagger}_{n\sigma{'}}(t_2)\hat{C}_{k'_\beta\sigma{'}}(t_{2})>\nonumber\\
&&-t_{k_\alpha m \sigma}t^{\ast}_{k'_\beta n \sigma{'}} <\hat{C}^{\dagger}_{k_\alpha\sigma}(t_{1})\hat{d}_{m\sigma}(t_{1})><\hat{d}^{\dagger}_{n\sigma{'}}(t_2)\hat{C}_{k'_\beta\sigma{'}}(t_{2})>\nonumber\\
&&-t^{\ast}_{k_\alpha m \sigma}t_{k'_\beta n \sigma{'}} <\hat{d}^{\dagger}_{m\sigma}(t_{1})\hat{C}_{k_\alpha\sigma}(t_{1})><\hat{C}^{\dagger}_{k'_\beta\sigma{'}}(t_{2})\hat{d}_{n\sigma{'}}(t_2)>]\nonumber.\\
\label{Curtimes1}
\end{eqnarray}

Using the Wick's theorem~\cite{GDMahan}, we have,
\begin{eqnarray}
<\hat{C}_{k_\alpha\sigma}^{\dagger}(t_{1})\hat{d}_{m\sigma}(t_1)\hat{C}_{k'_\beta\sigma{'}}^{\dagger}(t_{2})\hat{d}_{n\sigma{'}}(t_2)> \nonumber\\
=<\hat{C}_{k_\alpha\sigma}^{\dagger}(t_{1})\hat{d}_{m\sigma}(t_1)><\hat{C}_{k'_\beta\sigma{'}}^{\dagger}(t_{2})\hat{d}_{n\sigma{'}}(t_2)>\nonumber\\
+<\hat{C}_{k_\alpha\sigma}^{\dagger}(t_{1})\hat{d}_{n\sigma{'}}(t_2)><\hat{d}_{m\sigma}(t_1)\hat{C}_{k'_\beta\sigma{'}}^{\dagger}(t_2)>.
\label{wick theorem}
\end{eqnarray}
The shot noise can be expressed in terms of Green's function
\begin{eqnarray}
&&<\Delta\hat{I}_{\alpha\sigma}(t_{1})\Delta\hat{I}_{\beta\sigma{'}}(t_{2})>=\nonumber\\
&&-\frac{q^{2}}{\hbar^2}\sum_{kk{'}mn}[t_{k_\alpha m \sigma}t_{k'_\beta n \sigma{'}}G^{>}_{m\sigma k'_\beta\sigma{'}}G^{<}_{n\sigma{'}k_\alpha\sigma}\nonumber\\
&&+t^{\ast}_{k_\alpha m \sigma}t^{\ast}_{k'_\beta n \sigma{'}}G^{>}_{k_\alpha\sigma n\sigma{'}}G^{<}_{k'_\beta\sigma m\sigma}\nonumber\\
&&-t_{k_\alpha m \sigma}t^{\ast}_{k'_\beta n \sigma{'}}G^{>}_{m\sigma n\sigma{'}}G^{<}_{k'_\beta\sigma{'} k_\alpha\sigma}\nonumber\\
&&-t^{\ast}_{k_\alpha m \sigma}t_{k'_\beta n \sigma{'}}G^{>}_{k_\alpha\sigma k'_\beta\sigma{'}}G^{<}_{n\sigma{'}m\sigma }].
\label{Curtimes2}
\end{eqnarray}
From the Langreth theorem of analytic continuation, we have
\begin{eqnarray}
&&G^{<,>}_{m\sigma k'_\beta\sigma{'}}(t_1,t_2)\nonumber\\
&=& \sum_{p\sigma_{p}} \int dt [G^r_{m\sigma p\sigma_{p}}(t_1,t)
t^*_{k'_\beta p \sigma_p} g^{<,>}_{k'_\beta\sigma_{p}\sigma{'}}(t,t_2) \nonumber \\
&+& G^{<,>}_{m\sigma p\sigma_{p}}(t_1,t) t^*_{k'_\beta p \sigma_p} g^a_{k'_\beta\sigma_{p}\sigma{'}}(t,t_2)],
\label{X2conti1}
\end{eqnarray}
and
\begin{eqnarray}
&&G^{<,>}_{k'_\beta\sigma{'}m\sigma}(t_1,t_2)\nonumber\\
&=& \sum_{p\sigma_{p}} \int dt [g^{<,>}_{k'_\beta\sigma{'}\sigma_{p}}(t_1,t)
t_{k'_\beta p \sigma_p} G^a_{p\sigma_{p}m\sigma}(t_1,t) \nonumber \\
&+& g^r_{k'_\beta\sigma{'}\sigma_{p}}(t_1,t) t_{k'_\beta p\sigma_{p}} G^{<,>}_{p\sigma_{p}m\sigma}(t,t_2)],
\label{X2conti2}
\end{eqnarray}
as well as
\begin{eqnarray}
&&G^{<,>}_{k_\alpha\sigma k'_\beta\sigma{'}}(t_1,t_2) = g^{<,>}_{k'_\beta\sigma\sigma{'}}(t_1,t_2)
\delta_{kk'}\delta_{\alpha\beta} \nonumber \\
&+&\sum_{p\sigma_{p}} \int dt [G^r_{k_\alpha\sigma p\sigma_{p}}(t_1,t) t^*_{k'_\beta p \sigma_p}
g^{<,>}_{k'_\beta\sigma_{p}\sigma{'}}(t,t_2) \nonumber \\
&+& G^{<,>}_{k_\alpha\sigma p\sigma_{p}}(t_1,t) t^*_{k'_\beta p \sigma_p} g^a_{k'_\beta\sigma_{p}\sigma{'}}(t,t_2)],
\label{X2conti3}
\end{eqnarray}
The self-energy is given by
\begin{equation}
\Sigma^\gamma_{\alpha m\sigma n\sigma{'}}(t_1,t_2)= \sum_k t^*_{k_\alpha m \sigma}(t_1)
g^\gamma_{k_\alpha\sigma\sigma{'}}(t_1,t_2) t_{k_\alpha n \sigma{'}}(t_2).
\label{X2self}
\end{equation}

From the above equations, we can calculate the noise spectrum $S$ defined as follows~\cite{MButtikerPRB}:
\begin{equation}
\pi \delta(0) S_{\alpha\beta}^{\sigma\sigma{'}}=\int dt_{1}dt_{2}<\Delta\hat{I}_{\alpha\sigma}(t_{1})\Delta\hat{I}_{\beta\sigma{'}}(t_{2})>.
\label{ssgsg}
\end{equation}
Using Eqs.(\ref{X2conti1})-(\ref{X2self}), it is straightforward to write the noise spectrum as
\begin{equation}
S_{\alpha\beta}^{\sigma\sigma{'}}=\sum_{i=0}^{4} S_{i,\alpha\beta}^{\sigma\sigma{'}}.
\label{ssum}
\end{equation}
Here
\begin{equation}
\pi \delta(0) S_{0,\alpha\beta}^{\sigma\sigma{'}}=\frac{q^2}{\hbar^2} \delta_{\alpha\beta}Tr_{t}[{G^{>}_{\sigma\sigma{'}}}\Sigma^{<}_{\alpha,\sigma{'}\sigma}+{\Sigma^{>}_{\alpha,\sigma\sigma{'}}}G^{<}_{\sigma{'}\sigma}],
\label{S0}
\end{equation}
\begin{eqnarray}
\pi \delta(0) S_{1,\alpha\beta}^{\sigma\sigma{'}}=&&-\frac{q^2}{\hbar^2} Tr_{t}[ (G^{r}\Sigma^{>}_{\beta}+G^{>}\Sigma^{a}_{\beta})_{\sigma\sigma{'}}\nonumber\\
&&\times
(G^{r}\Sigma^{<}_{\alpha}+G^{<}\Sigma^{a}_{\alpha})_{\sigma{'}\sigma}],
\label{S1}
\end{eqnarray}
\begin{eqnarray}
\pi \delta(0) S_{2,\alpha\beta}^{\sigma\sigma{'}}=&&-\frac{q^2}{\hbar^2} Tr_{t}[ (\Sigma^{>}_{\alpha}G^{a}+\Sigma^{r}_{\alpha}G^{>})_{\sigma\sigma{'}}\nonumber\\
&&\times
(\Sigma^{<}_{\beta}G^{a}+\Sigma^{r}_{\beta}G^{<})_{\sigma{'}\sigma}],
\label{S2}
\end{eqnarray}
\begin{eqnarray}
\pi\delta(0) S_{3,\alpha\beta}^{\sigma\sigma{'}}=&&\frac{q^2}{\hbar^2} Tr_{t}[G^{>}_{\sigma\sigma{'}}(\Sigma^{r}_{\beta}G^{r}\Sigma^{<}_{\alpha}
+\Sigma^{<}_{\beta}G^{a}\Sigma^{a}_{\alpha}\nonumber\\
&&+\Sigma^{r}_{\beta}G^{<}\Sigma^{a}_{\alpha})_{\sigma{'}\sigma})],
\label{S3}
\end{eqnarray}
\begin{eqnarray}
\pi\delta(0) S_{4,\alpha\beta}^{\sigma\sigma{'}}=&&\frac{q^2}{\hbar^2} Tr_{t}[(\Sigma^{r}_{\alpha}G^{r}\Sigma^{>}_{\beta}+\Sigma^{>}_{\alpha}G^{a}\Sigma^{a}_{\beta}\nonumber\\
&&+\Sigma^{r}_{\alpha}G^{>}\Sigma^{a}_{\beta})_{\sigma\sigma{'}}G^{<}_{\sigma{'}\sigma}
].
\label{S4}
\end{eqnarray}
Since all Green's functions depend on double time indices, the trace $Tr_t$ is taken in the time domain.

To get the charge current noise spectrum, we combine equation (\ref{curnoise}) with
equation (\ref{ssgsg}) and perform Fourier transformation, the well known auto charge current noise
can be obtained
\begin{eqnarray}
S_{LL}=\frac{q^2}{\pi}\int dE (f_{L}-f_{R})^2 Tr[(1-T)T],
\label{auto-c}
\end{eqnarray}
here, $T=\Gamma_{L}G^{r}\Gamma_{R}G^{a}$ is the transmission matrix.
We can also get the cross charge current shot noise (along z direction):
\begin{eqnarray}
S_{LR}=-\frac{q^2}{\pi}\int dE (f_{L}-f_{R})^2 Tr[(1-T)T],
\label{auto-c1}
\end{eqnarray}
$S_{LL}+S_{LR}=0$ confirms the current conservation.

For spin current noise spectrum, the situation is very different. We combine equation
(\ref{spincurnoise}) with equation (\ref{ssgsg}), taking Fourier transformation, and using the relation
\begin{equation}
\sum_{\sigma\sigma{'}}\sigma\sigma{'}A_{\sigma\sigma{'}}B_{\sigma{'}\sigma}=Tr[A\sigma_{z}B\sigma_{z}],
\end{equation}
($\sigma_{z}$ is pauli matrix),  we can obtain the auto spin current shot noise (zero-temperature limit),
\begin{eqnarray}
S^{\sigma}_{LL}=\frac{\hbar^2}{4\pi}\int dE (f_{L}-f_{R})^2 Tr[\sigma_{z}T\sigma_{z}(1-T)].
\label{auto-sc}
\end{eqnarray}
Similarly, the cross spin current shot noise can be obtained:
\begin{eqnarray}
&&S^{\sigma}_{LR}=-\frac{\hbar^2}{4\pi}\int dE (f_{R}-f_{L})^{2}Tr\{[G^{r}
\Gamma_{R}G^{a}(\Sigma^{a}_{R}\sigma_{z}\nonumber\\
&&-\sigma_{z}\Sigma^{r}_{R})+G^{r}\Gamma_{R}\sigma_{z}](G^{r}
\Gamma_{L}\sigma_{z}+G^{r}\Gamma_{L}G^{a}(\Sigma_{L}^{a}\sigma_{z}\nonumber\\
&&-\sigma_{z}\Sigma_{L}^{r})\}.
\label{cross-sc}
\end{eqnarray}

Now, we derive the shot noise of spin transfer torque ${\cal S}(t_1,t_2)$,
\begin{eqnarray}
\hspace{-2.0cm}
{\cal S}(t_1,t_2)&&=-\frac{1}{4}<\sum_{k_{R_A}n}\sum_{\sigma_{A}\sigma_{A}^{'}}
(\hat{C}_{k_{R_A}\sigma_{A}}^{\dagger}\mathcal{R}_{\sigma_{A}\sigma_{A}^{'}}
\hat{d}_{n\sigma^{'}_{A}}t_{k_{R_A}n\sigma^{'}_A}
-t_{k_{R_A}n\sigma^{'}_{A}}^{*}\hat{d}_{n\sigma^{'}_{A}}^{\dagger}
\mathcal{R}_{\sigma_A\sigma_{A}^{'}}\hat{C}_{k_{R_A}\sigma_{A}})\nonumber\\
&&\sum_{k_{R_B}m}\sum_{\sigma_{B}\sigma_{B}^{'}}(\hat{C}_{k_{R_B}\sigma_{B}}^{\dagger}
\mathcal{R}_{\sigma_{B}\sigma_{B}^{'}}\hat{d}_{n\sigma^{'}_{B}}t_{k_{R_B}n\sigma^{'}_B}
-t_{k_{R_B}n\sigma^{'}_{B}}^{*}\hat{d}_{n\sigma^{'}_{B}}^{\dagger}\mathcal{R}_{\sigma_B\sigma_{B}^{'}}\hat{C}_{k_{R_{B}}\sigma_{B}})>
\end{eqnarray}
and
\begin{eqnarray}
\hspace{-2.5cm}
<\hat{\tau}_{R}(t)>=\frac{i}{2}<\sum_{k_{R_A}n}\sum_{\sigma_{A}\sigma_{A}^{'}}
(\hat{C}_{k_{R_A}\sigma_{A}}^{\dagger}\mathcal{R}_{\sigma_{A}\sigma_{A}^{'}}\hat{d}_{n\sigma^{'}_{A}}
t_{k_{R_A}n\sigma^{'}_A}-t_{k_{R_A}n\sigma^{'}_{A}}^{*}\hat{d}_{n\sigma^{'}_{A}}^{\dagger}
\mathcal{R}_{\sigma_A\sigma_{A}^{'}}\hat{C}_{k_{R_A}\sigma_{A}})>.
\end{eqnarray}
Similarly, we define the shot noise spectrum of spin transfer torque
like the shot noise of spin current, it can be written as
\begin{eqnarray}
\pi\delta(0)S^{\tau}&=&\int dt_{1}dt_{2} {\cal S}(t_1,t_2).
\end{eqnarray}
where ${\cal S}(t_1,t_2)$ is defined in equation (\ref{sttnoise}).

By using the Wick's theorem, and after Fourier transformation, we can obtain expression of $S^{\tau}$,
\begin{eqnarray}
\label{torquenoise}
S^{\tau}=&&\frac{\hbar^{2}}{4\pi}\int dE(f_{L}-f_{R})\{Tr[G^{r}\Gamma_{L}G^{a}\mathcal{R}\Gamma_{R}\mathcal{R}\nonumber\\
+&&G^{r}\Gamma_{L}G^{a}(\Sigma_{R}^{a}\mathcal{R}-\mathcal{R}
\Sigma_{R}^{r})G^{r}\Gamma_{L}G^{a}(\Sigma_{R}^{a}\mathcal{R}-\mathcal{R}\Sigma_{R}^{r})\nonumber\\
+&&G^{r}(i\mathcal{R}\Sigma_{R}^{r}-i\Sigma_{R}^{a}\mathcal{R}-\Gamma_{R}
\mathcal{R})G^{r}\Gamma_{L}G^{a}(\Sigma_{R}^{a}\mathcal{R}-\mathcal{R}\Sigma_{R}^{r})\nonumber\\
+&&(\Sigma_{R}^{a}\mathcal{R}-\mathcal{R}\Sigma_{R}^{r})G^{r}\Gamma_{L}G^{a}
(\mathcal{R}\Gamma_{R}-i\mathcal{R}\Sigma_{R}^{r}+i\Sigma_{R}^{a}\mathcal{R})G^{a}]\}.
\end{eqnarray}

\section{Shot noise of spin current and spin torque for MNM system}

In this paper, we consider a normal quantum dot connected
by two ferromagnetic leads (see figure 1) (MNM system). The magnetic moment of left lead is pointing to
the $z$-direction, while the moment of right lead is at an angle $\theta_R$ to the $z$-axis
in the $x-z$ plane. Hence the Hamiltonian of quantum dot
can be written as
\begin{equation}
H_{dot}=
\left(
  \begin{array}{ccc}
    \epsilon_0 & 0 \\
    0 & \epsilon_0\\
  \end{array}
\right).
\end{equation}

%Fig.2
\begin{figure}
\includegraphics[height=6cm,width=8cm]{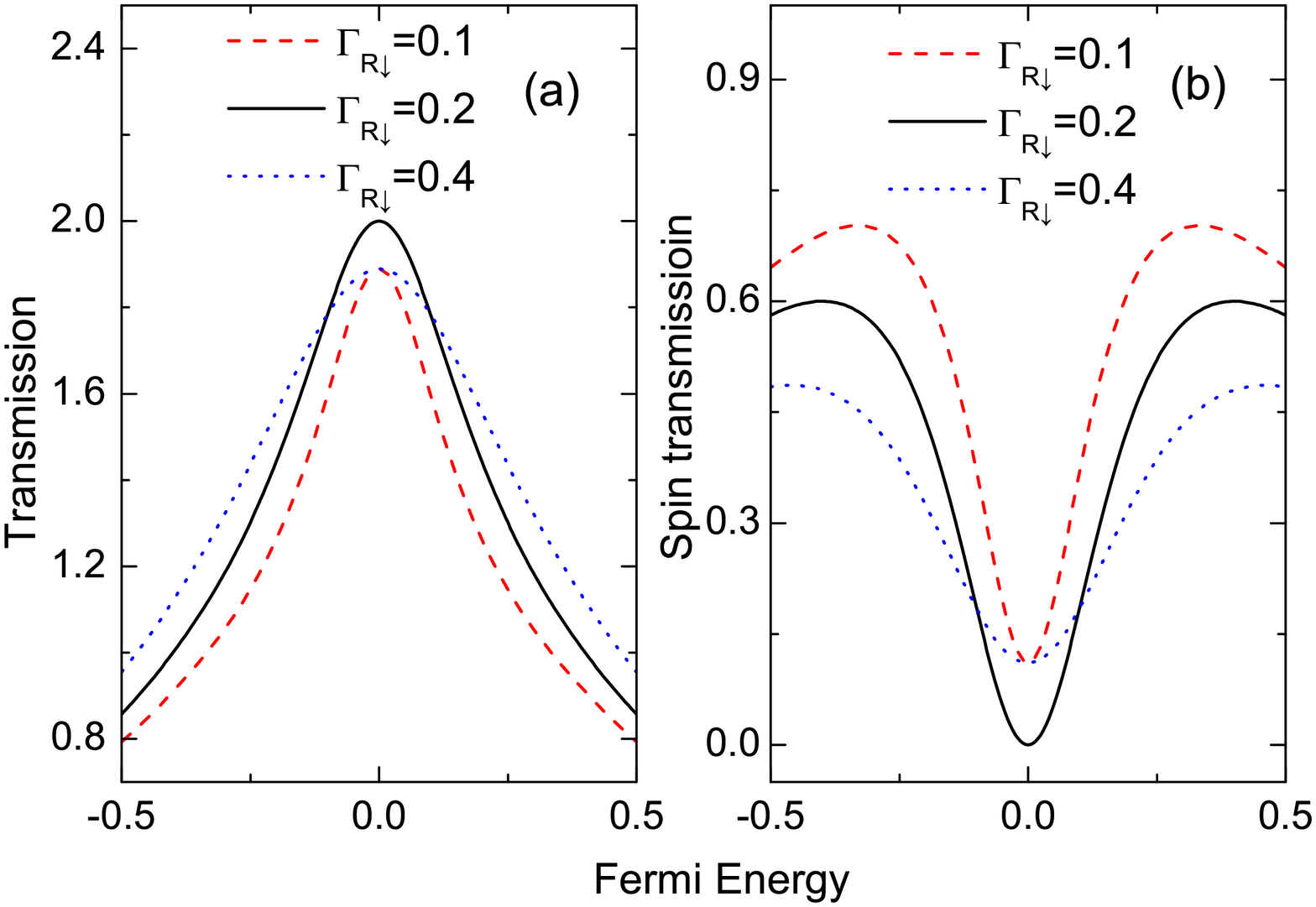}
\caption{(a)The charge transmission coefficient $(T_{\uparrow}+T_{\downarrow})$
and (b)the spin transmission coefficient $(T_{\uparrow}-T_{\downarrow})$
versus fermi energy when $\Gamma_{R\downarrow}=0.1$ (red dashed line),
$\Gamma_{R\downarrow}=0.2$ (black solid line), $\Gamma_{R\downarrow}=0.4$
(blue dotted line). The other parameters are $\theta_R=0$, $\epsilon=0$
$\Gamma_{L\downarrow}=0.2$, $\Gamma_{L\uparrow}=\Gamma_{R\uparrow}=0.8$.
The energy unit is eV.}
\label{Fig.2}
\end{figure}

Firstly, we set the direction of magnetization of the right lead be along the $z$ direction, i.e.,
let $\theta_R=0$ and calculate the charge and spin current according to Landauer-B$\ddot{u}$ttiker formula
\begin{eqnarray}
I_{c}=-\frac{q}{\hbar}\int\frac{dE}{2\pi}Tr[\hat{T}(E)](f_{L}-f_{R}),
\end{eqnarray}
and the expression of spin current is
\begin{eqnarray}
I_{s}=\frac{1}{2}\int\frac{dE}{2\pi}Tr[\sigma_{z}\hat{T}(E)](f_{L}-f_{R}).
\label{spincurrent}
\end{eqnarray}
The charge and spin transmission coefficients are depicted in figure 2.
In the calculation, we have chosen $\Gamma_{L\uparrow}=\Gamma_{R\uparrow}=0.8$eV
and fix the energy unit is to be eV. Let $\Gamma_{L\downarrow}=\Gamma_{R\downarrow}=0.2$eV
(Here, we let $\Gamma_{\alpha\uparrow}\neq\Gamma_{\alpha\downarrow}$ due to the presence
of ferromagnetic leads), we found that the charge transmission coefficient reaches two at the resonant energy level
$E=\epsilon_0$ of the quantum dot (solid line in the left panel of figure 2), while the
spin transmission coefficient is zero at resonant energy point
(solid line in the right panel of figure 2). For parallel situation ($\theta_R=0$) there is no
spin flip so that different spin channel can be treated separately. For a symmetric coupling from the lead,
both spin up and spin down electrons have complete transmission at the resonance. For total
charge current they add up together while for total spin current they cancel to each other.
When we break this symmetry and change $\Gamma_{R\downarrow}$ while keeping
$\Gamma_{L\downarrow}$ constant, the spin down transport
will be partially blocked, so the charge transmission coefficient
will decrease and the spin transmission coefficient will increase (see the dashed lines
and dotted lines in figure 2).

Figure 3 gives a comparison between the charge current and spin current
versus $\theta_R$ under the small bias voltage 0.05V.
From the figure, we find that for the symmetric coupling with $\Gamma_{L\uparrow}=\Gamma_{R\uparrow}$
and $\Gamma_{L\downarrow}=\Gamma_{R\downarrow}$, both charge current and spin current
decrease as $\theta_R$ increasing from zero to $\pi$ (see the solid lines in figure 3(a) and 3(b)).
But if we fix $\Gamma_{L\downarrow}$ and change $\Gamma_{R\downarrow}$, although the charge current still decreases when $\theta_R$ changes from zero to $\pi$, the spin current increases when $\Gamma_{R\downarrow} > \Gamma_{L\downarrow}$
and decreases when $\Gamma_{R\downarrow} < \Gamma_{L\downarrow}$ and changes sign at
$\theta_R=\pi$. To understand the behavior, we plot the spin-up current in the panel (c)
and spin-down current in the panel (d), respectively. One can clearly see that
spin up current always decreases with $\theta_R$ from zero to $\pi$, but spin down current
always increases though it is negative. So the competition between spin up and down channels determines how the total spin current varies with $\Gamma_{R\downarrow}$.
Another interesting result is that
at $\theta_R=\pi$, i.e., when the magnetic moments of the two leads are antiparallel, the spin down current
does not change when we change the $\Gamma_{R\downarrow}$
 (see figure \ref{Fig.4}(d)) while keeping other parameters the same.
In fact, when we change the $\Gamma_{R\downarrow}$ at $\theta_R=\pi$, we actually change the
right coupling line-width constant of spin up but not spin down due to $\Gamma_R(\pi)=R_{\alpha}(\pi)\left(
\begin{array}{ccc} \Gamma_{R\uparrow} & 0 \\  0 & \Gamma_{R\downarrow}\\ \end{array}
\right)R^{\dagger}_{\alpha}(\pi)=\left( \begin{array}{ccc} \Gamma_{R\downarrow} & 0 \\  0 & \Gamma_{R\uparrow}\\ \end{array}
\right) $. So we can find that at $\theta_R=\pi$, the
spin up current is different with different $\Gamma_{R\downarrow}$ but spin down keeps unchanged.

%Fig.3
\begin{figure}[tbp]
\includegraphics[height=6cm,width=8cm]{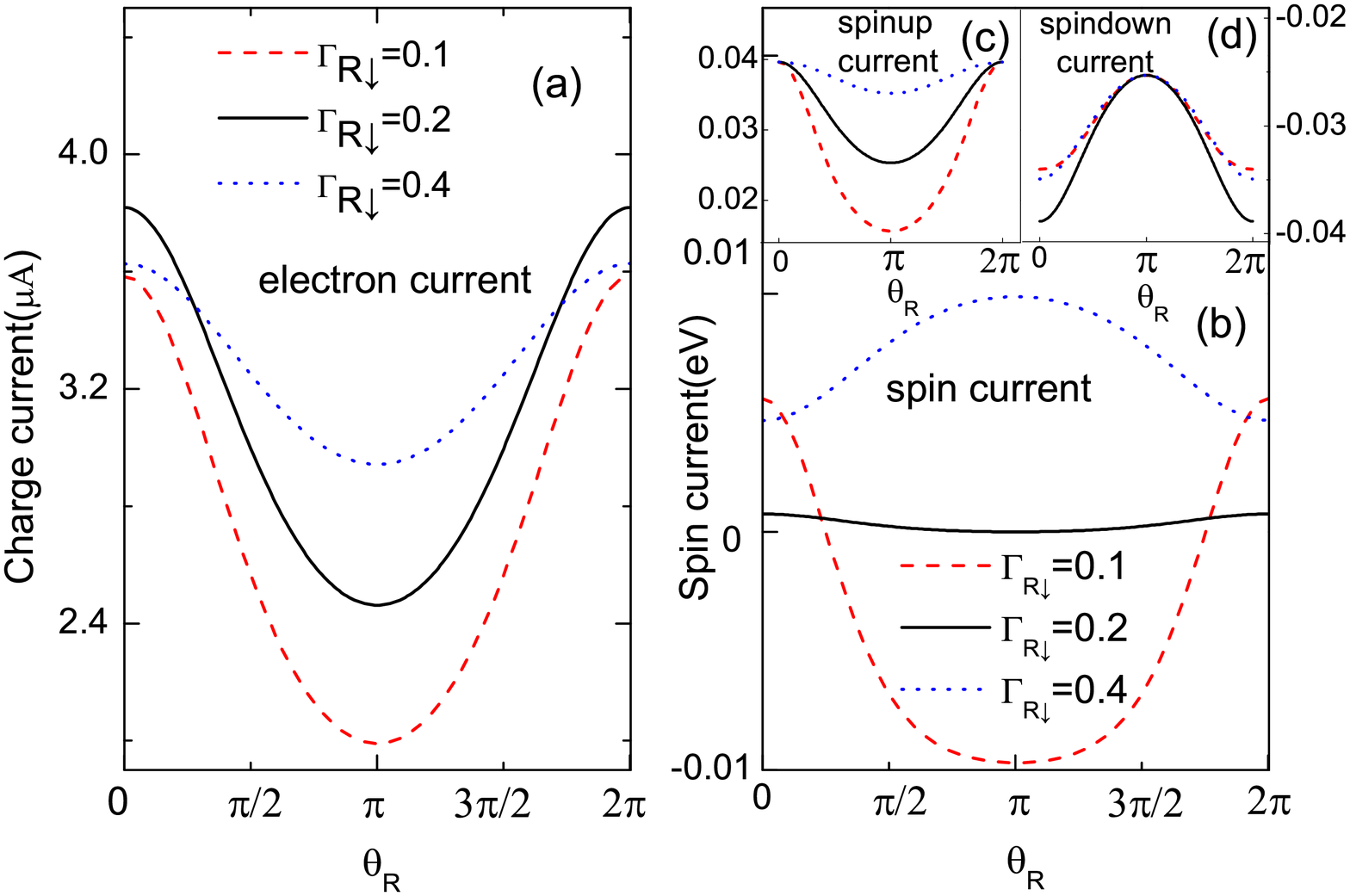}
\caption{The charge current (panel (a)), total spin current (panel (b)),
spin up current (panel (c)) and spin down current (panel (d))
versus $\theta_R$ for MNM system
at $\Gamma_{R\downarrow}=0.1$ (red dashed line),
$\Gamma_{R\downarrow}=0.2$ (black solid line),
$\Gamma_{R\downarrow}=0.4$ (blue dotted line),respectively.
The other parameters are $V_{bias}=0.05$V,
$\epsilon_0=0$, $\Gamma_{L\uparrow}=\Gamma_{R\uparrow}=0.8$,
$\Gamma_{L\downarrow}=0.2$. }
\label{Fig.3}
\end{figure}

To study the shot noise of spin current, we first examine the differential
shot noise spectrum versus bias voltage $V_{L}=V$ and $V_{R}=0$. At zero temperature,
they can be calculated from equations (\ref{auto-sc}) and (\ref{cross-sc})
(AC means auto-correlation and CC means cross-correlation)
\begin{eqnarray}
  N_{AC}=\frac{4\pi}{q\hbar^2} \frac{\partial S_{LL}^{\sigma_{z}}}{\partial V}
   =Tr[\sigma_{z}T(E)\sigma_{z}(1-T(E))]|_{E=qV}
\label{NAC}
\end{eqnarray}
and
\begin{eqnarray}
N_{CC}=\frac{4\pi}{q\hbar^2} \frac{\partial S_{LR}^{\sigma_{z}}}
 {\partial V}=-Tr\{[G^{r}\Gamma_{R}G^{a}(\Sigma^{a}_{R}\sigma_{z}-\sigma_{z}\Sigma^{r}_{R})+\nonumber\\
G^{r}\Gamma_{R}\sigma_{z}](G^{r}\Gamma_{L}\sigma_{z}+G^{r}
\Gamma_{L}G^{a}(\Sigma_{L}^{a}\sigma_{z}-\sigma_{z}\Sigma_{L}^{r}))\}|_{E=qV}.
\label{NCC}
\end{eqnarray}

For equation (\ref{NCC}), we see that if the direction of magnetic moments of both leads are parallel the off-diagonal matrix elements of all
the physical quantity including the linewidth function $\Gamma_{\alpha\sigma}$
are zero, so $\sigma_{z}$ commutes with other matrices in equation (\ref{NCC}).
Using this property and $G^{a}-G^{r}=iG^{r}\Gamma_{L}G^{a}+iG^{r}\Gamma_{R}G^{a}$, we find
\begin{eqnarray}
  N_{CC}=-Tr[\sigma_{z}T\sigma_{z}(I-T)]=-N_{AC}.
\end{eqnarray}

%Fig.4
\begin{figure}[tbp]
\includegraphics[height=7cm,width=9cm]{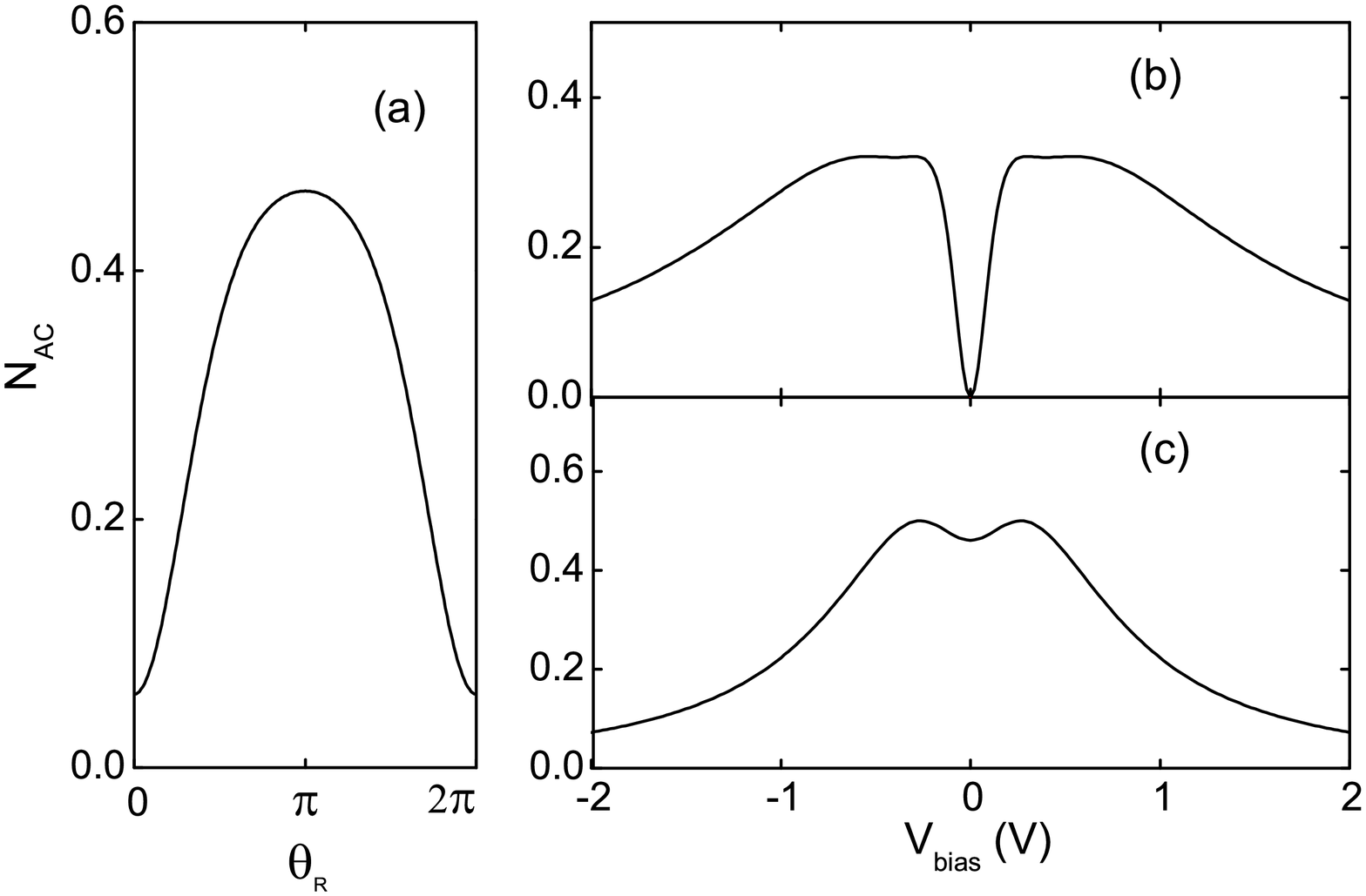}
\caption{MNM system. (a) $N_{AC}$ of
 spin current noise versus the angle $\theta_R$ with $V_{bias}=0.05$V; (b) $N_{AC}$ versus the bias voltage with two electrode
 magnetic moments parallel; (c) $N_{AC}$versus the bias voltage with two electrode magnetic moments antiparallel.
 The other parameters are $\epsilon_0=0$, and $\Gamma_{L\uparrow}=\Gamma_{R\uparrow}=0.8,\Gamma_{L\downarrow}=\Gamma_{R\downarrow}=0.2$.}
\label{Fig.4}
\end{figure}

Now we calculate $N_{AC}$ from equations (\ref{NAC}). The figure 4(a) gives $N_{AC}$ versus
$\theta_R$. One can find that the differential spin shot $N_{AC}$ is small for parallel situation and
reaches maximum when the magnetization of leads are antiparallel. We also plot differential spin shot noise versus bias voltage at parallel and antiparallel configurations in figure 4(b) and 4(c).  When the magnetization of
two lead are parallel, $N_{AC}$ increases abruptly with
the bias voltage and reaches a flat plateau between about $V_{bias}=(0.3,0.7)V$, then decreases gradually
upon further increasing bias voltage.  However, for antiparallel case, $N_{AC}$ starts at a large value compared with that of parallel case and increases a bit to a maximum value at $V_{bias}=\pm0.26V$.
For large bias voltage $V_{bias}$ $N_{AC}$ decreases and gradually approaches to zero. We have shown that $N_{AC}+N_{CC}=0$ in the case of parallel and anti-parallel situations. It is found that this relation is still valid when $\theta_R$ is not equal to $0$ or $\pi$.
In general, the relation $N_{AC}+N_{CC}$ is not satisfied. For instance, If we study a system MNMNM with three ferromagnetic layers or the MM interface where coupling matrix elements $\Gamma_{\sigma\bar\sigma} \neq 0$, one can
get $N_{AC}+N_{CC} \neq 0$.

Now we analyze the spin transfer torque and its auto-correlation function. From equations (\ref{spintorque}) and (\ref{torquenoise}), we calculate the derivative of spin transfer torque and its correlation function with respect to the bias voltage as follows:
\begin{eqnarray}
 T_{\tau}=\frac{2\pi}{q}\frac{\partial <\hat{\tau}_{R}>}{\partial V}=Tr[G^{r}\Gamma_{L}G^{a}(i\Sigma_{R}^{a}\mathcal{R}-i\mathcal{R}\Sigma_{R}^{r})]|_{E=qV}, \label{ttau}
\end{eqnarray}
and
\begin{eqnarray}
N_{\tau}&=&\frac{4\pi}{q\hbar^2}\frac{\partial S^{\tau}}{\partial V}=Tr[G^{r}\Gamma_{L}G^{a}(\Sigma_{R}^{a}\mathcal{R}-\mathcal{R}
\Sigma_{R}^{r})G^{r}\Gamma_{L}G^{a}(\Sigma_{R}^{a}\mathcal{R}-\mathcal{R}\Sigma_{R}^{r})\nonumber\\
&+&G^{r}\Gamma_{L}G^{a}\mathcal{R}\Gamma_{R}\mathcal{R}\nonumber\\
&+&G^{r}(i\mathcal{R}\Sigma_{R}^{r}-i\Sigma_{R}^{a}\mathcal{R}-\Gamma_{R}
\mathcal{R})G^{r}\Gamma_{L}G^{a}(\Sigma_{R}^{a}\mathcal{R}-\mathcal{R}\Sigma_{R}^{r})\nonumber\\
&+&(\Sigma_{R}^{a}\mathcal{R}-\mathcal{R}\Sigma_{R}^{r})G^{r}\Gamma_{L}G^{a}
(\mathcal{R}\Gamma_{R}-i\mathcal{R}\Sigma_{R}^{r}+i\Sigma_{R}^{a}\mathcal{R})G^{a}]\}|_{E=qV}.
\label{Ntau}
\end{eqnarray}

\begin{figure}[tbp]
\includegraphics[height=6cm,width=8cm]{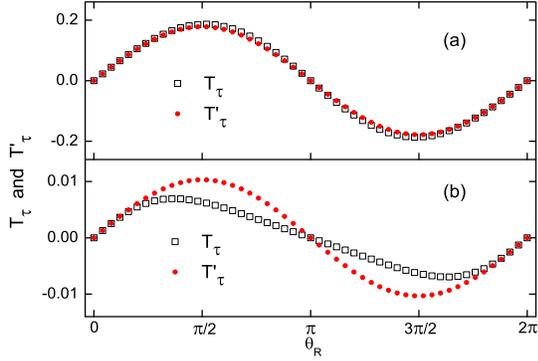}
\caption{$T_{\tau}$ and $T'_{\tau}$ versus $\theta_R$ with different bias voltages
$V=0.01V$ (panel (a)) and $V=0.97V$ (panel (b)). The other parameters are
$\epsilon_0=1.0$, $\Gamma_{L\uparrow}=\Gamma_{R\uparrow}=0.8,
\Gamma_{L\downarrow}=\Gamma_{R\downarrow}=0.2$.
}
\label{Fig.5}
\end{figure}

Since most of calculations for the spin transfer torque were obtained using the
formula\cite{IoannisTheodonis,Guo1,Guo2} $\tau_0=\frac{I^s(\pi)-I^s(0)}{2}\sin\theta$, we also calculate
$T'_{\tau}=\frac{2\pi}{q}\frac{\partial \tau_0}{\partial V}$ for comparison. In figure
\ref{Fig.5}, we plot $T_{\tau}$ and $T'_{\tau}$ versus $\theta_R$.
When the bias voltage is tuned far away from the resonant point $\epsilon_0$ (figure 5(a)), the profile
of $T_{\tau}$ versus $\theta_R$ obeys $\sin\theta_R$ function. This gives very good agreement
with $T'_{\tau}$ which is expected since $T'_{\tau}$ was derived for a non-resonant tunneling system. When the system is near resonance, however, $T_{\tau}$ deviates away from the sinusoidal dependence\cite{Sugang1,Sugang2}. This behavior can be understood as follows. When we set
$\Gamma_{L\uparrow}=\Gamma_{R\uparrow}=\Gamma_{\uparrow}$
and $\Gamma_{L\downarrow}=\Gamma_{R\downarrow}=\Gamma_{\downarrow}$,
equation (\ref{ttau}) can be simplified as
\begin{eqnarray}
\hspace{-1.0cm}
T_\tau=\frac{1}{2}\frac{(qV-\epsilon_0)^2 (\Gamma_{\uparrow}^{2}-\Gamma_{\downarrow}^{2})\sin\theta}
{(qV-\epsilon_0)^2(\Gamma_{\uparrow}+\Gamma_{\downarrow})^2+[(qV-\epsilon_0)^2-\Gamma_{\uparrow}\Gamma_{\downarrow}-\frac{1}{4}
(\Gamma_{\uparrow}-\Gamma_{\downarrow})^2\sin^2\frac{\theta}{2}]^2}.
\label{Tau11}
\end{eqnarray}
We examine the denominator of this equation. Since $\Gamma_{\uparrow} \ne \Gamma_{\downarrow}$, it is clear that near the resonance $qV \sim \epsilon_0$, the term $\sin^2(\theta/2)$ in the denominator cannot be neglected so that $T_{\tau}$ in the upper panel of figure (\ref{Fig.5}) is not the $\sin\theta_R$ dependence. But when $|{qV-\epsilon_0}|$ is large enough so that the term $\sin^2(\theta/2)$ is small compared with the term $(qV-\epsilon_0)^2$,we obtain $T_{\tau} \approx T'_{\tau}$.
Actually, we can derive $T'_{\tau}$ by differentiating $I^s(\pi)$ and $I^s(0)$ according to equation (\ref{spincurrent}) and obtain
\begin{eqnarray}
T'_\tau=\frac{1}{2}\frac{(qV-\epsilon_0)^2 (\Gamma_{\uparrow}^{2}-\Gamma_{\downarrow}^{2})\sin\theta}
{[(qV-\epsilon_0)^2+\Gamma_{\uparrow}^2][(qV-\epsilon_0)^2+\Gamma_{\downarrow}^2]}.
\label{Tau22}
\end{eqnarray}
One can easily find that if we neglect the term $\sin^2(\theta/2)$ in the denominator of $T_{\tau}$, $T_{\tau}$ will equal $T'_{\tau}$.

%Fig.5

%Fig.6
\begin{figure}[tbp]
\includegraphics[height=7cm,width=8cm]{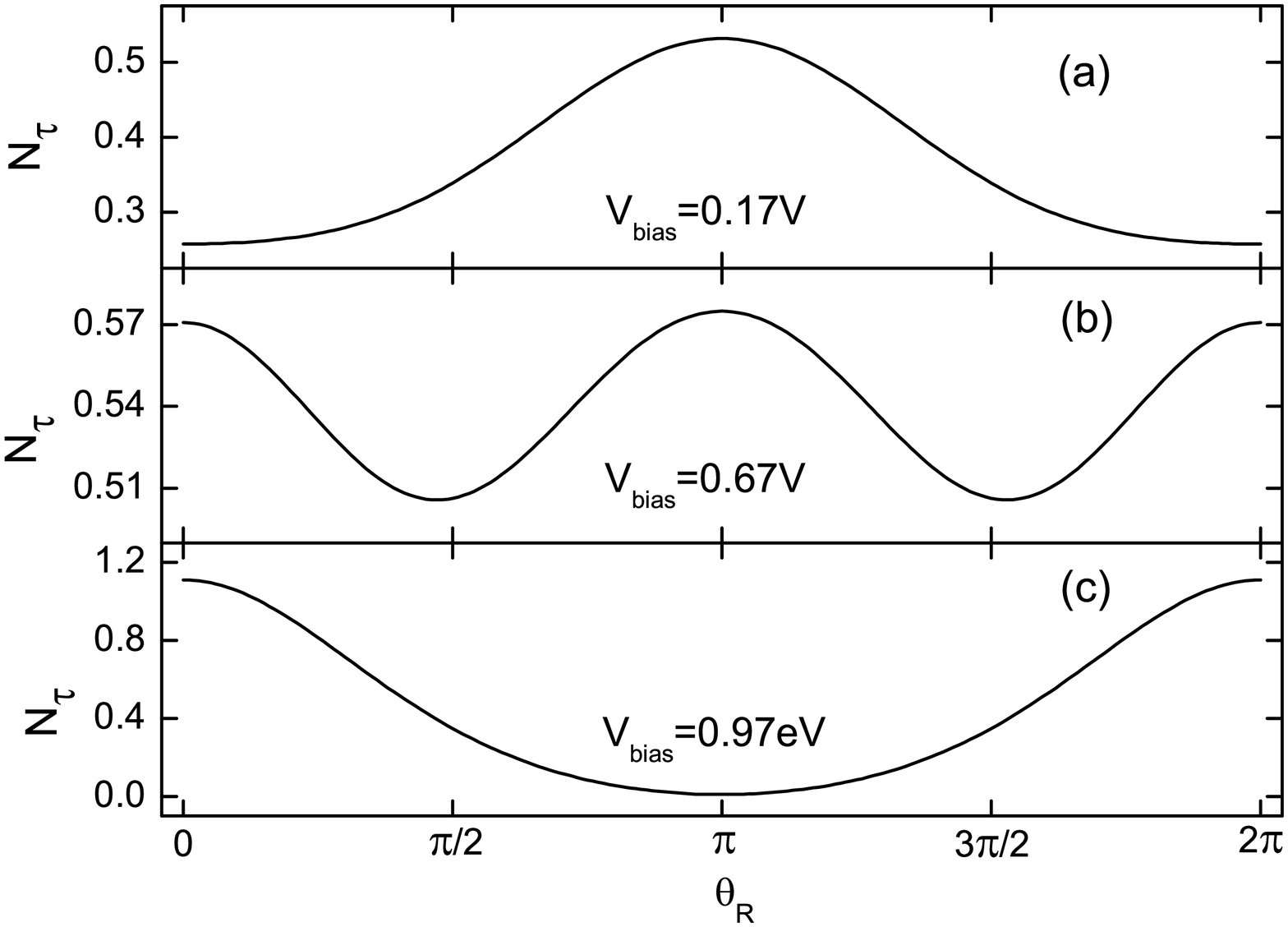}
\caption{$N_{\tau}$ versus the angle $\theta_R$ with the bias voltage
$V_{bias}=0.17V$ (panel (a)), $V_{bias}=0.67V$ (panel (b)) and $V_{bias}=0.97V$ (panel (c)).
The other parameters are $\epsilon_0=1.0$, and
$\Gamma_{L\uparrow}=\Gamma_{R\uparrow}=0.8,\Gamma_{L\downarrow}=\Gamma_{R\downarrow}=0.2$.
}
\label{Fig.6}
\end{figure}

Finally, we calculated derivative of the noise spectrum of spin transfer torque with respect to the bias voltage
by equation (\ref{Ntau}). From figure (\ref{Fig.6}), we see that $N_\tau$ as a function of $\theta_R$ gives very different behaviors depending on whether it is near resonance or far away from that. When the bias voltage is close to $\epsilon_0/q$, i.e., when the system is near resonance (figure 6(c)), $N_\tau$ is a concave function of $\theta_R$ which
is very large at $\theta_R=0$ but close to zero at $\theta_R=\pi$. However, when the system is far away from the resonance, $N_\tau$ is is convex function of $\theta_R$ that is small at $\theta_R=0$ but large at $\theta_R=\pi$ (see figure (\ref{Fig.6})(a)). In the intermediate range of bias voltage, the differential noise spectrum of spin transfer torque behaves like $\sin(2\theta_R)$ (see figure (\ref{Fig.6})(b)). When we change $\Gamma_{R\downarrow}$ and keep the other parameters the same, we found that the noise spectrum of spin transfer torque is very sensitive to $\Gamma_{R\downarrow}$ when $\theta_R$ is near zero.

\section{Conclusions}
In conclusion, based on the Green's function approach, the spin current and spin noise of quantum dot coupled by two ferromagnetic leads were investigated. The spin auto-correlation function is always positive while the spin cross-correlation noise is negative definite. Due to the existence of the spin flip, the sum of them can be non-zero for systems with three ferromagnetic layers, i.e, $S_{LL}^{\sigma_z}+S_{LR}^{\sigma_z} \neq 0$. As a result, both the spin auto-correlation noise and spin cross-correlation noise are needed to characterize the shot noise of spin current. The spin transfer torque and its noise spectrum were also investigated. For a system with a resonant level, the differential spin transfer torque was found to be proportional to $\sin\theta$ far away from the resonance where $\theta$ is the angle between magnetization of two ferromagnetic leads. Near the resonance, however, a non-sinusoidal $\theta$ dependence was found. The noise spectrum of spin transfer torque is found to be a concave function of $\theta$ near the resonance and becomes a convex function far away from the resonance. The noise spectrum of spin transfer torque was found to be very sensitive to the system parameters and might be used to characterize the electron spin transport properties.

\ack{
We gratefully acknowledge support by the grant from the National Natural Science
Foundation of China with Grant No.10947018(Y.J. Yu) and No.11074171(Y.D. Wei), No. 11274364 (Q.F. Sun), and a GRF grant from HKSAR (HKU 705611P) (J. Wang).

\end{document}